%% file: main.tex
\documentclass[conference]{IEEEtran}

\usepackage{url}
\usepackage{xurl}
\usepackage{cite}
\usepackage{amsmath,amssymb,amsfonts}
\usepackage{algorithmic}
\usepackage{graphicx}
\usepackage{textcomp}
\usepackage{xcolor}
\usepackage{hyperref}
\usepackage{stfloats}
\usepackage{booktabs}
\usepackage{microtype}

\def\BibTeX{{\rm B\kern-.05em{\sc i\kern-.025em b}\kern-.08em
  T\kern-.1667em\lower.7ex\hbox{E}\kern-.125emX}}
\begin{document}

\title{Gradient-Based Learning of Parametric Engine Sound Representations for Real-Time Resynthesis and Tuning on Embedded Systems
}

\author{
\IEEEauthorblockN{
Robin Doerfler,
Matthieu Kuntz,
Clemens Zimmer
}
\IEEEauthorblockA{
Impulse Audio Lab GmbH\\
Munich, Germany\\
robin.doerfler@impulse-audio-lab.com,\\
matthieu.kuntz@impulse-audio-lab.com,\\
clemens.zimmer@impulse-audio-lab.com
}
}

\maketitle

\begin{abstract}
Engine order enhancement is central in automotive sound design, where selective harmonics are synthesized to shape perceptual qualities such as sportiness, refinedness, or power. This paper investigates a neural network-based approach to combustion engine sound modeling that extends conventional engine order analysis and enhancement by deriving synthesis parameters from audio data with machine learning and incorporating stochastic components into the synthesis framework. The system parameterizes engine sounds as a compact representation capturing per-order and broadband timbral variation across the full RPM-torque operating range, while remaining manually tunable and compatible with established automotive audio frameworks. The approach leverages gradient-based optimization and analysis-by-synthesis through an end-to-end differentiable implementation. The resulting synthesis parameter set is directly transferable to conventional DSP implementations for deployment on embedded targets. Spectral metrics and listening tests\footnotemark confirm high reconstruction fidelity, and integration into an established automotive audio development platform (\textit{EVx Suite}) demonstrates technical feasibility on deployment-ready embedded systems.
\end{abstract}


\footnotetext{Audio examples are available online\\
~\url{https://rdoerfler.github.io/eone-model-page/}}

\input{introduction}
\input{methods}
\input{results}
\input{discussion}
\input{summary}

\bibliographystyle{IEEEtran}
\bibliography{refs}

\end{document}

%% file: introduction.tex
\section{Introduction}
\label{sec:introduction}

A central application of Active Sound Design (ASD) in the automotive audio domain is Engine Order Enhancement (EOE), in which individual harmonics of the engine sound are amplified to enhance perceptual qualities such as sportiness, refinedness, or power \cite{kwon2018sportiness, moonActiveSoundDesign2020}, thereby supporting the overall driving experience \cite{bisping1995emotional} and brand identity \cite{changResearchBrandSound2017}. 

Conventional EOE systems synthesize a set of harmonics with amplitudes defined as functions of vehicle parameters, such as engine speed, torque, or pedal position \cite{boddenComprehensiveAutomotiveActive2014, kim2024SmartphonebasedASD}. These harmonics, referred to as \emph{engine orders}, correspond to sinusoidal components at integer and half-integer multiples of the crankshaft rotation frequency. An initial parameter set of such amplitude functions is commonly derived by Engine Order Analysis (EOA) of recordings \cite{herlufsen1999characteristics, saavedraAccurateAssessmentComputed2006, nielsen2017parametric} and subsequent energy evaluation at the harmonics \cite{wangVehicleNoiseVibration2010, doerflerAnalysisDrivenProceduralGeneration2026} for any given engine operating-state. Those parameters are then integrated and tuned by an engineer in the target vehicle, effectively combining a synthesis starting point grounded in real acoustic behavior with fine-grained adjustment during application.

However, in recent years a discernible shift toward higher fidelity and more realistic engine sound modeling \cite{dupreAnalysisSynthesisEngine2023} has appeared. With the growing adoption of combustion engine sound simulation for electric vehicles \cite{Yu2024FrequencyShiftEV, Zhou2025ShiftSoundQuality}, ASD applications increasingly extend beyond enhancement of individual engine orders, requiring more comprehensive acoustic replication.

While alternative synthesis paradigms such as sample-based playback \cite{heitbrinkDesignDrivingSimulation2007, chenSynthesisingSoundCar2021}, hybrid analysis-synthesis \cite{banSynthesisCarNoise2002, liRealTimeAutomotiveEngine2024} or fully procedural methods \cite{baldanPhysicallyInformedCar2015, dupreAnalysisSynthesisEngine2023} exist, they offer limited support for the per-harmonic, per-operating-state adjustments required during in-vehicle tuning. Conversely, recent neural audio synthesis methods have demonstrated promising results in replicating complete engine sounds through end-to-end learning \cite{lundbergDataDrivenProceduralAudio2020, lobatoMotor2SynthLeveragingDifferentiable2025, doerflerPhysicsInformedNeuralEngine2026}, but are not primarily designed for seamless integration into existing EOE frameworks or efficient implementation on embedded systems.

Given the established workflows and versatile application potential of parametric engine-order frameworks, they remain among the most practical foundations for tuning-oriented, deployment-ready ASD systems --- despite the insufficient acoustic realism of purely harmonic representations and the tuning complexity that grows with number of operating-state dependent synthesis components \cite{boddenComprehensiveAutomotiveActive2014}.

Extending such frameworks toward comprehensive engine sound replication requires incorporating the full harmonic spectrum of engine orders and broadband noise components, while maintaining a parametrization compact enough to remain manually tunable --- ideally reducing amplitude controls for each synthesis component to a single gain function over engine speed, expressed in revolutions per minute (RPM), and torque respectively, the dominant determinants of engine acoustic output.

However, such a compact, principled representation cannot be derived analytically: no single analysis frame contains sufficient context to resolve whether energy at a given frequency originates from a harmonic or noise, nor how a required gain adjustment should be distributed between the RPM and torque dimensions. The only principled resolution is to treat parametrization as an optimization problem, letting sequential, multi-frame data resolve what no analytical method can.

We propose the \textit{Engine-Order and Noise Extraction (EONE)} model, a neural network that learns parametric engine sound representations through differentiable analysis-by-synthesis \cite{engelNeuralAudioSynthesis2017}: network parameters directly encoding gain functions over RPM and torque --- equivalents to conventional lookup tables for both engine orders and noise bands --- are optimized by matching synthesis outputs to target recordings. This approach sidesteps DSP-based extraction entirely, as the network discovers appropriate parametrization strategies from pattern recognition in data.

To enable fast convergence, robustness to confounding noise, and generalization from limited target recordings, a timbre encoder-decoder is first pretrained on a large corpus of diverse engine recordings. At deployment, only a lightweight model with the gain functions as its exclusive learnable parameters is fine-tuned to the target material, initialized from the pretrained latent timbre representations. 

The proposed method differs from prior work in three key aspects: it jointly models tonal and stochastic noise components within a unified EOE framework; it derives a compact, globally valid parametrization across the full RPM-torque space without exhaustive grid sampling; and its parameters map directly onto conventional EOE DSP algorithms, ensuring one-to-one translation to embedded deployment.

Spectral-similarity evaluation between target signals and reconstructions, as well as perceptual validation through an listening test, demonstrate sufficient reconstruction fidelity for comprehensive engine sound modeling despite compression into a compact parametric representation of gain functions per harmonic and noise band. Crucially, integration within an established automotive audio framework (\textit{EVx Suite}) demonstrates technical feasibility on deployment-ready embedded systems, while providing a user interface for direct parameter inspection and further manual refinement.

%% file: methods.tex
\section{Methods}

\subsection{Factorized RPM-Torque Gain Curve Representation}
\label{sec:parametric_representation}
The objective is to parameterize engine sounds as a compact representation that captures per-order and broadband timbral variation across the full RPM-torque operating range, remains manually tunable, and is compatible with established EOE synthesis frameworks.

\subsubsection{Harmonic and Noise Band Amplitude Factorization}
Engine sounds are modeled as a sum of harmonic engine orders with frequency trajectories at integer multiples of the firing frequency ($f_0 = \text{RPM}/120$), and broadband noise components, where per-harmonic and per-band amplitudes fully determine the acoustic output at any given RPM-torque state.

A full per-component 2D amplitude map over RPM and torque would yield the most accurate approximation, but obtaining such dense parameter grids requires comprehensive data coverage and becomes intractable for manual tuning. Factorizing into independent 1D functions over RPM and torque reduces memory overhead and aligns naturally with traditional EOE workflows.

However, independent 1D functions are under-constrained: a desired gain adjustment at any RPM-torque operating point can be absorbed by either function, with persistent effect across the full range of the other axis. For instance, on a given harmonic, a -3\,dB offset applied at 6000\,RPM will attenuate that harmonic across the entire torque range, regardless of torque state, and vice versa.

The ambiguity in this factorization cannot be resolved analytically, the optimal distribution of gain between RPM and torque functions must be determined jointly from data. This applies equally to the separation of harmonic and noise band contributions: spectral energy at any harmonic frequency in a single analysis frame may originate from either component, and their individual contributions can only be reliably estimated with multi-frame context, motivating sequence-level, gradient-based optimization by synthesis.

Among several factorization strategies explored, including an additive log-domain combination and a variant constraining the torque function to a scalar offset, a simple multiplicative combination of RPM and torque gain functions yielded the best reconstruction results under identical training conditions (same initialization and number of epochs), and was adopted for all experiments.

\subsubsection{Engine Order and Noise Extraction by Resynthesis}
\label{sec:analysis_by_synthesis}
Instead of traditional engine order extraction methods, we adopt an analysis-by-synthesis approach, determining synthesis parameters directly by comparing the synthesized output to target acoustic signals. This offers inherent robustness to FFT windowing artifacts, spectral leakage, and bin spilling from non-stationary measurements. Rather than relying on classical DSP criteria such as peak-picking strategies or energy integration thresholds --- where analysis parameter choices directly determine the extracted values --- gradient flow through the full synthesis path optimizes gain parameters against reconstruction targets, decoupling parameter values from any specific analysis configuration.

The differentiable synthesis implementation can be mirrored directly on the conventional DSP running on an embedded system, ensuring that learned parametric representations translate without modification to efficient low-level implementations for real-time resynthesis on target hardware. 

\subsection{EONE Model Architecture}
\label{sec:model_architecture}
\begin{figure*}[htbp]
\centering
\includegraphics[width=\textwidth]{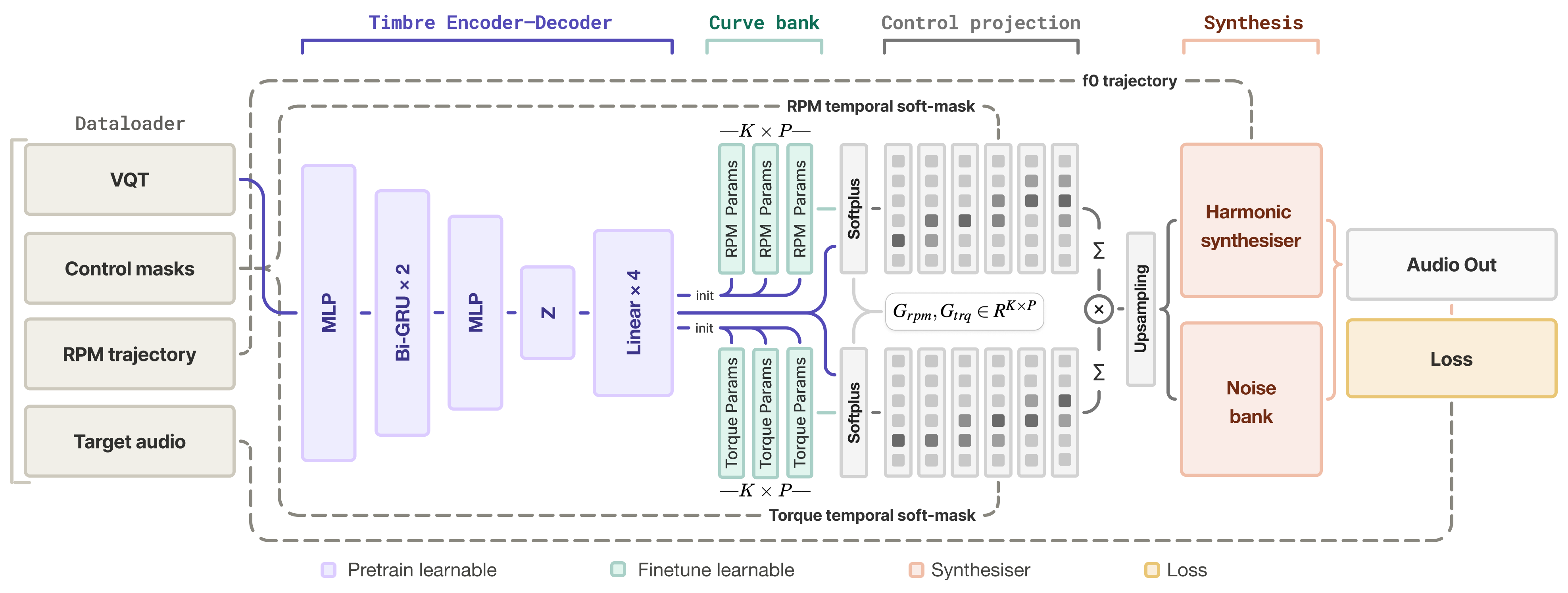}
\caption[EONE model overview]{%
Input audio is provided as VQT spectrograms and encoded to a compact latent timbre representation $z$, before being decoded into two sets of Softplus-activated gain curves $G_\text{rpm}, G_\text{trq} \in \mathbb{R}^{K \times P}$, forming the shared parametrization for end-to-end training and direct DSP export at inference. RPM and torque gain curves are each projected onto their respective temporal soft-masks to yield time-varying amplitude envelopes, which are multiplied element-wise and upsampled to audio rate. The resulting amplitudes drive a differentiable harmonic synthesizer with $f_0$ derived from the RPM trajectory, and an ERB noise bank. During pretraining, encoder and decoder are jointly optimized; during finetuning, the curve bank parameters --- initialized from the pretrained decoder output --- are optimized directly. Training minimizes a combined MRSTFT and harmonic loss against the target audio.
}
\label{fig:architecture_overview}
\end{figure*}

The model is organized into four stages as shown in Figure~\ref{fig:architecture_overview}: the Timbre Encoder-Decoder, the Curve Bank, the Control Projection, and the Synthesis. The following sections detail each stage, training regimes and hyperparameters are described in Section~\ref{sec:training}.

\subsubsection{Timbre Encoder-Decoder}
\label{sec:timbre_encoder}
The Timbre Encoder-Decoder builds a structured latent space of engine timbre from which gain curves can be reliably inferred for engines not seen during pretraining.

The encoder operates on Variable-Q Transform (VQT) spectrograms \cite{schorkhuberMatlabToolboxEfficient2014, cheukNnAudioOntheFlyGPU2020}, which provide logarithmically uniform frequency resolution with a constant number of bins per octave. This ensures that harmonic patterns of engine orders shift as a rigid group across the frequency axis as RPM varies, preserving their relative spacing and making timbral structures consistent regardless of operating speed.

Raw VQT spectrograms, however, vary considerably across engines, microphone positions, and recording conditions. A frame projection compresses each VQT frame to a common feature dimension, a bidirectional GRU integrates spectral patterns across time, and its output is averaged across frames to yield a single fixed-size summary vector --- collapsing temporal information into a global timbral descriptor consistent with the time-invariant, operating-state-dependent gain curves. A bottleneck MLP projects this descriptor to $z \in \mathbb{R}^{32}$, organizing engine timbres by acoustic similarity in a compact latent space.

Because this space is trained on a large diversity of engine recordings, a new engine timbre maps to a meaningful coordinate surrounded by acoustically similar examples --- gain curves decoded from that coordinate reflect learned timbral structures rather than arbitrary initialization, and generalize beyond the control coverage of the target recording.

Four linear decoder heads map $z$ to RPM and torque gain parameters for harmonic and noise components. These parameters are passed through a Softplus activation function (a smooth approximation of ReLU) to form the initial Curve Bank $G_\text{rpm}, G_\text{trq} \in \mathbb{R}^{K \times P}$. At fine-tuning, encoder and decoder weights are frozen and only the initialized curve bank parameters are carried forward for direct optimization.

\subsubsection{From Gain Curves to Sequence}
\label{sec:gain_curves_to_sequence}
The model is constrained to predict synthesis parameters compatible with additive-noise synthesis frameworks. While the pretraining encoder leverages bidirectional GRU for temporal context during training, inference eventually runs in real-time against instantaneous RPM and torque updates from the vehicle CAN bus. Decoded parameters must therefore be functions of instantaneous control state rather than temporal sequences, preventing recurrent decoding at inference time.

We address this by parameterizing timbre as two sets of gain curves $G_\text{rpm}, G_\text{trq} \in \mathbb{R}^{K \times P}$, where $K$ is the number of synthesis components, consisting of $K_h$ harmonics and $K_b$ noise bands, and $P$ is the number of learnable points in the curve. In our experiments we set $P = 40$, corresponding to a 250~RPM and 25~Nm resolution for RPM and torque respectively.

Rather than indexing these curves directly --- which would be non-differentiable and incompatible with gradient-based optimization --- we project them onto time-varying soft control masks.

This linear interpolation scheme ensures that gradients flow into curve points to the exact degree of their activation at each observed control state, enabling localized updates proportional to control coverage in the training data.

For a sequence of $T$ control states, the RPM mask $M_\text{rpm} \in \mathbb{R}^{T \times P}$ encodes each control state as a linear interpolation between the two nearest of $P$ uniformly-spaced bins in the normalized control range ($\text{RPM} \in [0, 1]$ and $\text{torque} \in [-1, 1]$).

The per-timestep gain activation for each component is then:
\begin{equation}
    A_\text{rpm} = M_\text{rpm}\, G_\text{rpm}^\top \in \mathbb{R}^{T \times K}
    \label{eq:control_projection}
\end{equation}
and equivalently for torque, yielding $A_\text{trq}$. This operation is illustrated in the Control Projection stage of Figure~\ref{fig:architecture_overview}.

The gain curves $G_\text{rpm}$ and $G_\text{trq}$ are the direct 
optimization targets: static tables of $K \times P$ parameters, independent of sequence length. 

Nevertheless, the training  operates on sequences: spectral losses require sufficient temporal context to resolve frequency content, and gradient coverage across the curve points depends on the control variation observed within each training chunk. The soft mask projection generates these sequences from static curves via a single matrix multiplication: the differentiable counterpart to lookup-table interpolation in conventional DSP implementations. At inference, the projection collapses exactly to a table lookup with Softplus-activated curve parameters read out directly by control state on the embedded target without any neural network at runtime.

\subsubsection{Differentiable Synthesis}
Following our analysis-by-synthesis approach, we implement a fully differentiable synthesizer for end-to-end gradient-based optimization, designed to be directly translatable to conventional DSP implementations for deployment on the embedded target. The operation at an audio rate of 16\,kHz ensures that the $f_0$ trajectories exactly match the target, avoiding interpolation-induced deviations even at lower model frame rates.

\paragraph{Harmonic Synthesizer}
\label{subsec:harmonic_synth}
Engine orders are synthesized following traditional additive synthesis, with a time-varying $f_0$ trajectory derived from the RPM signal scaling a set of harmonic multipliers $m \in \{0.5, 1.0, 1.5, \ldots, m_{K_h}\}$ (we use \(K_h = 64\) in this work) to per-order frequency trajectories, onto which the predicted amplitudes are applied. Phase is accumulated via cumulative summation of the instantaneous normalized angular frequency:
\begin{equation}
    \phi_k(t) = \phi_0 + \sum_{\tau=0}^{t} \frac{2\pi m_k \cdot f_0(\tau)}{f_s}
\end{equation}
with a randomized initial offset $\phi_0 \sim \mathcal{U}(0, 2\pi)$ per chunk to ensure magnitude optimization remains independent of phase relationships. The instantaneous amplitude $a_k(t) = [A_\text{rpm}]_{t,k} \cdot [A_\text{trq}]_{t,k}$ is the product of the RPM and torque gain activations at time step $t$ for harmonic $k$, following the control projection of Section~\ref{sec:gain_curves_to_sequence}.

The synthesized harmonic component is then:
\begin{equation}
    y_h(t) = \sum_{k=1}^{K_h} a_k(t) \cdot \cos(\phi_k(t))
\end{equation}
This mirrors the behavior of interpolated lookup-based oscillators used in the embedded implementation, while remaining fully differentiable with respect to the predicted amplitudes for end-to-end training.

\paragraph{Filtered Noisebank}
\label{sec:noisebank}
Stochastic components of the engine sound are synthesized using a bank of filtered noise bands. Each band is obtained by multiplying a cosine-shaped magnitude filter with an independent random-phase spectrum and applying the inverse real-valued FFT, thereby exploiting the implicit periodicity of the inverse DFT: each reconstructed frame constitutes exactly one period of a band-limited periodic signal. Computing the noise bands over the full audio chunk thus yields perfectly looping per-band signals that require no further processing to serve as sample-based oscillators in the resynthesis engine.

A bank of $K_b$ ERB-spaced cosine filters (here $K_b=64$) partitions the spectrum into critical bands, mimicking the frequency-dependent processing of the inner ear. Each filter $F_b$ is centered at $c_b$ with bandwidth $\Delta_b$ in ERB-rate:
\begin{equation}
F_b(f) = \cos\!\left(\frac{\mathcal{E}(f) - c_b}{\Delta_b}\,\pi\right)
\end{equation}
where $\mathcal{E}(f) = 21.4\log_{10}(1 + 0.00437f)$ is the ERB-rate transform \cite{glasberg1990erb}. Multiplying each filter response by an independent random-phase spectrum and applying the IRFFT yields $B$ stationary stochastic bands $\{\eta_b(t)\}$. 
Just as in the harmonic synthesis, the instantaneous per-band amplitude $a_b(t) = [A_\text{rpm}]_{t,b} \cdot [A_\text{trq}]_{t,b}$ is the element-wise product of the two learned gain curves. 

The synthesized noise component is:
\begin{equation}
    y_n(t) = \sum_{b=1}^{K_b} a_b(t) \cdot \eta_b(t)
\end{equation}
The final synthesizer output is the sum of both components:
\begin{equation}
    y(t) = y_h(t) + y_n(t)
\end{equation}

\subsection{Real-time Resynthesis on Embedded Systems}
The embedded resynthesis mirrors the differentiable synthesizer structure exactly, ensuring that the parameters optimized during training translate directly to the target system. Harmonic components are implemented using memory-efficient interpolated lookup-based oscillators, and noise bands are derived from the pre-rendered per-band signals obtained via the IRFFT procedure described above.

The full synthesis parametrization is derived from the model after fine-tuning on the target engine recording set, yielding a compact set of gain curves that serve as the sole interface between the learned representation and the DSP layer. This parametric form enables convenient integration into low-level embedded systems and conventional ASD frameworks, while remaining interpretable and directly editable by engineers.

We implement the resynthesis backend within \textit{EVx Suite}, an automotive sound development platform providing both a desktop tuning tool and DSP deployment across embedded hardware targets. Engineers can derive synthesis parameters from their own engine recordings using the integrated pretrained model, inspect and manually refine the resulting gain curves through the tuning interface, and deploy the final parametrization to the vehicle, supporting established ASD workflows while leveraging automatic data-driven parametrization.

\subsection{Training and Optimization}
\label{sec:training}
The training procedure is divided into two stages. First, a timbre encoder-decoder is pretrained on a large and diverse corpus of engine recordings to learn generalizable engine timbre representations. Second, the gain curve parameters are fine-tuned directly on the target material, with the pretrained encoder-decoder serving solely as an initialization source. This separation ensures that the computational burden of learning broad engine acoustics is absorbed during pretraining, while fine-tuning remains lightweight and targeted.

\subsubsection{Pretraining the Timbre Encoder-Decoder}
We train the Timbre Encoder-Decoder on a large corpus of engine recordings spanning controlled dynamometer measurements in a semi-anechoic chamber and  field recordings from real driving scenarios. The dataset comprises recordings from diverse vehicles, engine and exhaust system configurations, and multiple microphone positions, from close-miked engine placements, to in-cabin and exterior measurements. We further augment the corpus with the \textit{Procedural Engine Sounds} Dataset \cite{doerflerProceduralEngineSounds2025}, a large-scale collection of procedurally generated engine audio free of confounding background noise. 

All recordings include time-aligned RPM and torque annotations embedded directly in the multichannel audio stream. The first channels carry the engine audio signal, while the final two channels encode the control parameters: engine rotational speed normalized to $[0, 1]$ via $\text{RPM} \times 10^{-4}$, and engine torque normalized to $[-1, 1]$ via $\text{Nm}\times 10^{-3}$.

The audio data is processed in batches of 16, each containing 65{,}536-sample mono chunks (4.096 seconds at 16\,kHz) extracted with 50\% overlap. While these audio chunks serve as reconstruction targets, the model input features are derived from a sequence of frames with a hop length of 256 samples (64\,ms intervals), comprising harmonic spectra computed via the VQT and per-frame averaged control signals converted into time-varying soft masks as described in Section~\ref{sec:gain_curves_to_sequence}. The model is trained for 200 epochs using the AdamW optimizer with one-cycle cosine annealing.

\subsubsection{Fine-Tuning Gain Curve Bank}
At fine-tuning, the pretrained Timbre Encoder-Decoder weights are frozen, and a single set of gain curve parameters becomes the sole optimization target --- the same parameters that are indexed directly by instantaneous control state to drive synthesis on the embedded target, without any neural network evaluation at runtime. Fine-tuning thus optimizes exactly the deployed representation, with the encoder-decoder serving only as initialization.

The curve parameters are initialized by averaging the gain curves predicted by the pretrained model over a warm-up set, providing a plausible starting point grounded in patterns learned during pretraining. They are then optimized jointly from parallel evaluations across multiple RPM-torque trajectories within each batch, allowing updates to be informed by a broad range of operating states. This initialization is particularly important when target recordings do not provide full RPM-torque coverage, and extrapolation relies on patterns established during pretraining. Combined with the substantial reduction in learnable parameters relative to the full encoder-decoder, this warm start enables rapid adaptation from only a few epochs of limited material.

To encourage well-behaved solutions, a curvature regularization term based on second-order derivatives along the gain curve points is applied, promoting smooth gain functions and reducing overfitting to local peaks.

\subsubsection{Loss Function Design}
We employ multi-resolution short-time Fourier Transform (STFT) loss with FFT sizes ranging from $32{,}768$ down to $32$ samples (75\% overlap, Hann window), capturing spectral structure across multiple resolutions. The loss combines spectral convergence, linear magnitude, log-magnitude, and spectral energy terms with equal weighting and scale-invariant normalization across all resolutions.

An additional harmonic loss supervises frame-wise energy along predicted engine-order harmonics, inspired by Campbell diagrams from rotating machinery analysis. Energies are computed from magnitude spectrograms masked along harmonic tracks derived from instantaneous RPM, using high spectral resolution (FFT size of $65{,}536$, window size of $16{,}384$, and hop size of $256$) to minimize spectral leakage and isolate harmonic regions at low fundamental frequencies.

\subsection{Evaluation}
We evaluate synthesizer outputs against target recordings using model-generated parameter sets, on a test set unseen during training. The target recordings were captured without systematic operating-point sampling in a semi-anechoic dynamometer environment, and vary in size and engine operation mode coverage, making the evaluation representative of practical real-world data conditions.

\subsubsection{Reconstruction Quality Assessment}
We evaluate the reconstruction quality using the well-established log spectral distance (LSD). For each target/synthesis sample pair, we compute their spectrograms $S_1$ and $S_2$ using a STFT with a window and FFT size of 4096, and a hop size of 512. We discard the bins below 10\,Hz (first order harmonic for some idle engine recordings), in order to avoid low-frequency noise influencing the error metric. The RMS log spectral distance is computed on each time frame on the frequency range $[10, 8000]$\,Hz, and further averaged across time:
\begin{equation}
    \mathrm{LSD} = \frac{1}{T} \sum_t \sqrt{\frac{1}{F} \sum_f \left( \log S_1(f,t) - \log S_2(f,t) \right)^2}.
\end{equation}
\subsubsection{Perceptual Evaluation}
A listening experiment with twelve participants was conducted to assess the perceptual quality of the EONE approach. Eleven reported normal hearing; one participant had diagnosed hearing loss and wore a hearing aid on the right ear. Six participants were qualified as ``expert'' listeners based on professional experience with combustion engine sounds; the remaining six were qualified as ``naive'' listeners.

Participants rated the realism of each presented stimulus by answering ``How realistic does this stimulus sound?'' on a scale from 0 to 100 using a horizontal slider. After each playback, the slider was reset to a random position. Each presented stimulus was faded in and out using a squared-cosine window of 20\,ms, and its level was roved by a random offset drawn from a uniform distribution between $-3$ and $3$\,dB to discourage loudness-based comparisons across trials.

Stimuli were drawn from one of three conditions: target recordings (REC), their respective EONE reconstructions, or truncated reconstructions using only the 36 first harmonics of the EONE approach~\cite{zellerHandbuchFahrzeugakustikGrundlagen2012}, referred to as EOE. The EOE condition is expected to perform worse than the others, as it omits broadband components. It was included as a representative baseline for currently used automotive sound design methods.

A total of 243 trials (27 stimuli $\times$ 3 conditions $\times$ 3 repetitions) were presented in random order, split into three runs of approximately 10 minutes each with short breaks in between. No training or dummy trials were included, allowing participants to rate stimuli based on their prior experience.

%% file: results.tex
\section{Results}

\subsection{Reconstruction Quality}
The LSD was computed on the 27 target-synthesis sample pairs presented to the listeners. Averaging across these samples yields a mean LSD of 1.13 with a standard deviation of 0.19. Following \cite{rabiner1993fundamentals}, this corresponds to a mean LSD of 4.9\,dB and a standard deviation of 0.8\,dB. 

Most of the error arises from narrow-band deviations between target and reproduction, likely caused by confounding noises such as dynamometer, wind, or tire sounds that are not strictly RPM-torque coupled and thus cannot be reconstructed from the control signals. Computing the LSD on 1/12\textsuperscript{th}-octave band-smoothed spectrograms reduces the LSD to 2.8\,dB.

\subsection{Perceptual Evaluation}
\begin{figure}[h]
\centering
\includegraphics[width=1.0\columnwidth]{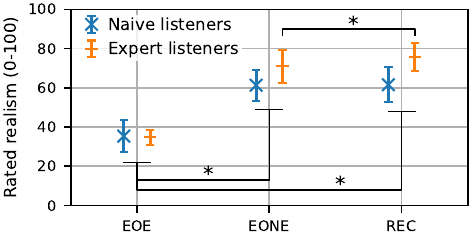}
\caption{%
Mean realism ratings for naive and expert listeners across EOE, EONE, and REC. Asterisks ($*$) indicate significant differences between bracketed groups: EOE vs.\ EONE, EOE vs.\ REC, and EONE vs.\ REC (experts only).
}
\label{fig:perceptual_results}
\end{figure}

Figure~\ref{fig:perceptual_results} shows the mean ratings grouped by listener experience and stimulus type. For statistical analysis, the raw ratings were averaged across the three repetitions, yielding one mean rating per participant per stimulus. The EOE stimulus was rated at 35 for both groups, although we observed a smaller standard deviation in the expert listeners. Naive listeners rated both EONE and REC with a score of 61, whereas expert listeners rated EONE and REC with 71 and 76 respectively.

\begin{figure}[t]
\centering
\includegraphics[width=0.928\columnwidth]{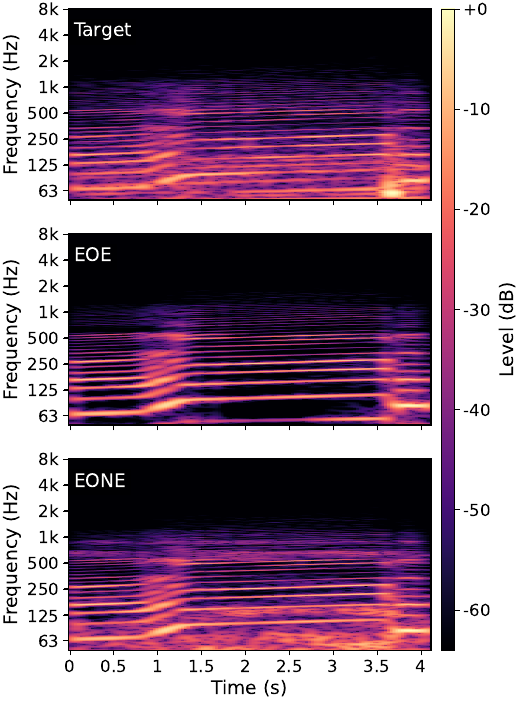}
\caption{%
Comparison of target recording and corresponding EOE and EONE predictions for a representative listening test sample. Inclusion of stochastic components yields spectral characteristics closer to the target, consistent with perceptual evaluation results.
Spectrograms computed using STFT with a Blackman–Harris window of 8192 samples, FFT size 32768, and hop size 512 samples. Color indicates magnitude level in dB.
}
\label{fig:listening_example_comparison}
\end{figure}

Figure~\ref{fig:listening_example_comparison} compares a representative example from the listening test and showing the effect of the stochastic components on the reconstructed spectrogram.

A two-way repeated-measures ANOVA was conducted with stimulus type as the within-subject factor and listener experience as the between-subject factor. Both main effects of stimulus type (\mbox{$F(2,20) = 72.35$, $p < 0.001$}) and listener experience (\mbox{$F(1,10) = 5.88$, $p = 0.036$}) were significant, whereas their interaction was not (\mbox{$F(2,20) = 2.94$, $p = 0.076$}). Pairwise comparisons using Bonferroni-corrected two-sided \textit{t}-tests revealed significant differences between EOE and EONE, and between EOE and REC, but no significant difference between EONE and REC ($p = 0.13$). We additionally tested for differences between EONE and REC only for the expert listener group using a paired \textit{t}-test and found a significant difference ($p<0.03$).

%% file: discussion.tex
\section{Discussion}
\label{sec:discussion}

The perceptual evaluation demonstrates that EONE achieves realism statistically comparable to real recordings for naive listeners, and while expert listeners could distinguish the two, EONE rated significantly above conventional EOE for both groups. The incorporation of stochastic components alongside harmonic synthesis is the primary driver of this improvement, extending timbral coverage beyond what conventional EOE methods can achieve. Spectral reconstruction quality further supports this, with residual error likely attributable to non-RPM-torque-coupled confounding noise rather than synthesis fidelity. Together these results indicate that for ASD engineers, the proposed system offers an efficient and principled alternative to systematic engine sound sampling and DSP-based order analysis --- one that requires only limited recordings, produces a convenient two-curve parametrization compatible with established per-order tuning, and deploys directly on embedded hardware without runtime neural network dependency.

The engine sound is modeled solely as a function of RPM and torque, which prevents representation of non-deterministic or occasional sound events, such as backfiring or turbo noises. During pretraining with temporal awareness on large datasets, these rare events are effectively ignored, but during fine-tuning on limited data they can leave persistent artifacts that appear tied to specific RPM–torque pairs regardless of temporal context. More broadly, while reducing the control space to just two dimensions is generally effective, it introduces deviations for physically unlikely operating combinations: the model is biased toward optimizing for frequently observed trajectories, so outlier conditions inherently diverge more strongly from real acoustic behavior. A higher-dimensional or full matrix representation could mitigate this, with trade-offs discussed in Section~\ref{sec:parametric_representation}.

%% file: summary.tex
\section{Summary}

This paper presented EONE, a neural network-based comprehensive combustion engine sound modeling approach extending conventional engine order analysis and enhancement methods by deriving synthesis parameters directly from recordings, and incorporating stochastic components into the synthesis framework. The system uses an analysis-by-synthesis approach to parametrize engine timbre as independent RPM and torque gain curves over harmonic and stochastic components, learned through a differentiable synthesizer constrained to produce representations directly transferable to conventional DSP implementations on automotive embedded hardware.

A pretrained Timbre Encoder-Decoder initializes the gain curves from a compact latent representation of engine timbre, enabling generalization beyond the control coverage of limited target recordings and reducing fine-tuning requirements to a small dataset. The full pipeline is integrated into the \textit{EVx Suite} automotive sound development platform, where engineers can derive synthesis parameters from their own recordings, inspect and refine the resulting gain curves, and deploy directly to the vehicle.

Spectral similarity evaluation on unseen recordings confirms high reconstruction fidelity despite the compact representation, supporting standalone applications such as electric vehicle interior sound design. A listening test demonstrates perceptual realism comparable to real engine noise and improvements over conventional engine order enhancement methods, which lack stochastic modeling capability.